\documentclass[letterpaper,11pt,leqno]{article}
\usepackage{paper}
\usepackage{circledsteps}
\begin{document}

\title{Emotions are Recognized Patterns of Cognitive Operations in Closed Control Loops}

\author{Yue Jin
%
\thanks{Yue Jin. Nokia Bell Labs France. Email: yue.1.jin@nokia-bell-labs.com.}}

\begin{titlepage}
\maketitle

Emotions play a crucial role in human life. The research community has proposed many theories of emotions without reaching consensus. Artificial Intelligence research uses emotion theories in the construction of artificial minds. It can also bring fresh perspective to emotion theories. Based on my observations on Soar cognitive architecture, I propose in this paper that the mind pursues goals with closed control loops and emotions are recognized patterns of cognitive operations in the control loops. The mind uses these operations to detect and eliminate the deviations between the targets of its goals and the outcomes of its actions. I map the patterns of cognitive operations to emotions. I present different complexities in the control loops. I show how my theory connects to existing theories and goes beyond them. I provide my answers to the five questions of emotion psychology. I show how psychological problems may arise when the mind fails to adapt the control loops. My theory advances the building of a more accurate theory of emotions that deepens the understanding of human minds and accelerates the construction of artificial minds.

\end{titlepage}

\section{Introduction}

Emotions play a crucial role in human life. They seem to compel us to pursue our goals, enhance or hinder our learning and decision-making, strengthen or weaken our communications and connections. Many theories have been proposed on what emotions are and how they are elicited. \cite{10.1098/rstb.2009.0141} groups the theories into three major models: Basic, Constructivist and Appraisal emotion theories. Basic emotion theories propose a set of basic emotions triggered by specific affect programs. Constructivist emotion theories propose that ``An emotion is your brain’s creation of what your bodily sensations mean, in relation to what is going on around you.'' (\cite{barrett2017emotions}). Appraisal emotion theories propose that an organism appraises a stimulus event against a standard set of criteria and the patterns of the appraisal outcomes are categorized as emotions. The research community is yet to reach consensus.

Artificial Intelligence community, Affect Computing in particular, builds emotion system for cognitive architectures based on the theories of emotion. According to \cite{larue2018emotion}, Soar-Emote uses emotional appraisals to frame information before it is processed by the cognitive system; Sigma models low-level appraisals as architectural self-reflection; ACT-R/$\Phi$ combines a model of a physiological substrate and the primary-process affect theory to augment the ACT-R cognitive architecture. 

The exchange between Artificial Intelligence and emotion theories goes both way. Artificial intelligence can bring fresh and complementary perspective to emotion theories. Based on my observations on Soar cognitive architecture, I propose in this paper that the mind pursues goals with closed control loops and emotions are recognized patterns of cognitive operations in the control loops. I first use Soar cognitive architecture to demonstrate in Section \ref{sec:soar} how surprise, nervousness and relief arise even when the mind only performs logical reasoning. I then make the case for minds in general on an extended list of emotions and present three complexities in the operations of control loops and emotions in Section \ref{sec:cog}. I show how my proposition connects to existing  theories of emotions and answer the five questions on theories of emotions in Section \ref{sec:literature}. I present my conjecture on how psychological problems may arise in Section \ref{sec:problem}. I conclude in Section \ref{sec:discuss} with a discussion on additional implications of my theory.

\section{Surprise, Nervousness and Relief in Soar} \label{sec:soar}

Soar is a long standing Cognitive Architecture (\cite{laird2022introduction}). It has gone through multiple versions with gradually added modules and capabilities since its conception in the early 80s. One of its capabilities is responding to incomplete knowledge. Whenever Soar doesn't have sufficient knowledge to select or apply an action, it labels this as an impasse. To resolve the impasse, it creates a substate with memories and processing powers and searches for a new action in the substate with all the processing available in Soar, including planning, retrievals from semantic and episodic memory and interaction with the external environment. Once a new action is found, Soar invokes a learning mechanism called ``Chunking'' to compile the processing in the substate into a new rule that it stores for future use.

\begin{figure}[t]
\centering
\includegraphics[width=0.8\columnwidth]{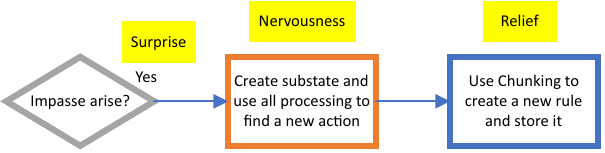} 
\caption{Emotions in Soar when Impasse Arises}
\label{fig:emotions_soar}
\end{figure}

We can clearly see that Soar goes through several stages and performs a distinctive cognitive operation in each stage when it encounters and resolves an impasse. When it encounters an impasse, this deviates from its goal. The detection of the deviation can be recognized as ``Surprise''. It then creates a substate and searches for a new action. It uses additional memory and processing power. This operation can be recognized as ``Nervousness''. Even though it's not yet a feature in Soar, it can invoke more parallel processing when it has less time to find a new action. The more parallel processing there is, the more intense the ``Nervousness'' is. When a new action is found, Soar invokes chunking to learn a new rule and stores it for future use. This operation can be recognized as ``Relief''. Figure \ref{fig:emotions_soar} shows the cognitive operations and corresponding emotions. For Soar to have emotions, it doesn't need a module to create the emotions; it needs a module to recognize the patterns of these cognitive operations.

\section{Closed Control Loops and Emotions} \label{sec:cog}

\subsection{Basics of Closed Control Loops and Emotions} \label{sec:basics}

\begin{figure}[t]
\centering
\includegraphics[width=\columnwidth]{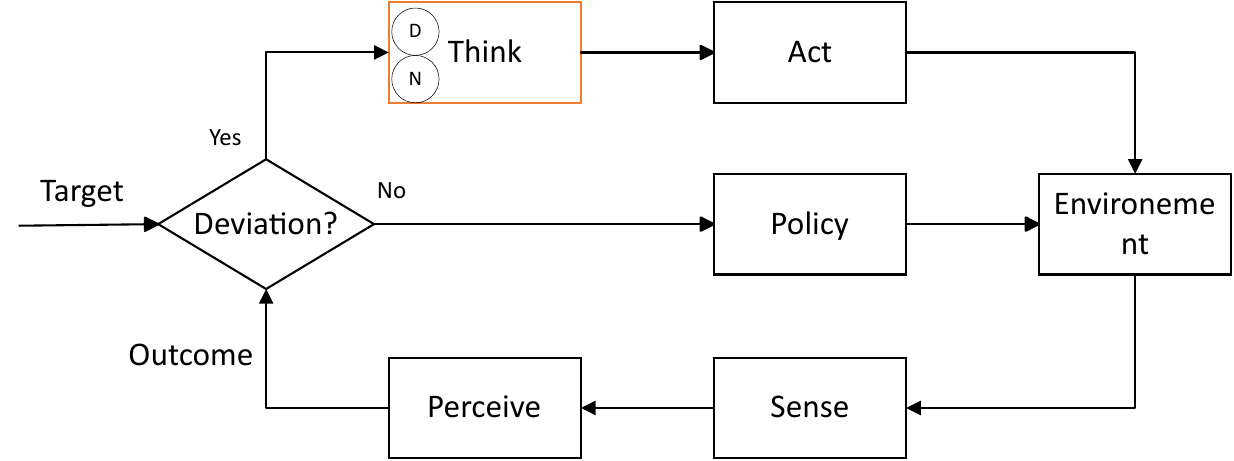} 
\caption{The mind pursues a goal with a closed control loop}
\label{fig:closed_loop}
\end{figure}

A mind pursues goals through the mechanism of closed control loops. I depict one such closed control loop in Figure \ref{fig:closed_loop}. It has a desired value for the goal, which I refer to as a ``target''. It has a set of actions to reach the target for different external and internal conditions, which I refer to as a ``policy''. The control loop executes the policy, receives external and internal sensory data and compiles the data into perception of the action outcome. It compares the outcome with the target. If the outcome conforms to the target, it keeps applying the policy. If the outcome deviates from the target, it signals the event to the mind. The mind allocates cognitive resources to assemble a ``Think-Act'' branch in the control loop and learns to eliminate the deviation. It appraises whether the outcome is better or worse than the target and processes the deviation accordingly. 

If the deviation is worse, the mind attempts to find a new action that restores the target. It has 2 layers in the ``Think-Act'' branch: an inner layer where it attempts to find a new action to reach the target and an outer layer where it lowers the target when the inner layer fails to find an action. It allocates a learning budget, which I represent approximately with the numbers of trials in the inner and outer layers ($N$ and $D$ in Figure \ref{fig:closed_loop} respectively) for ease of exposition. 

In each trial of the inner layer, it proposes a new action in ``Think'' through all available processing including planning, searching memory, random composition, etc. It executes the action and perceives its outcome by compiling sensory data. It compares the outcome to the target. If the outcome still deviates from the target, it starts another trial. It stops the inner layer when either the outcome conforms to the target or the number of trials $N$ is exhausted. In the former case, it updates the policy with the new action for future use. It then dissembles the ``Think-Act'' branch. In the latter case, it lowers the target and restarts the inner layer with the number of trials $N$. It stops the outer layer when either a new action is found for a lower target or the number of trials $D$ is exhausted. In the former case, it updates the policy and the target for future use and dissembles the ``Think-Act'' branch. In the latter case, it terminates its attempts and deletes the control loop. 

\begin{figure}[t]
\centering
\includegraphics[width=\columnwidth]{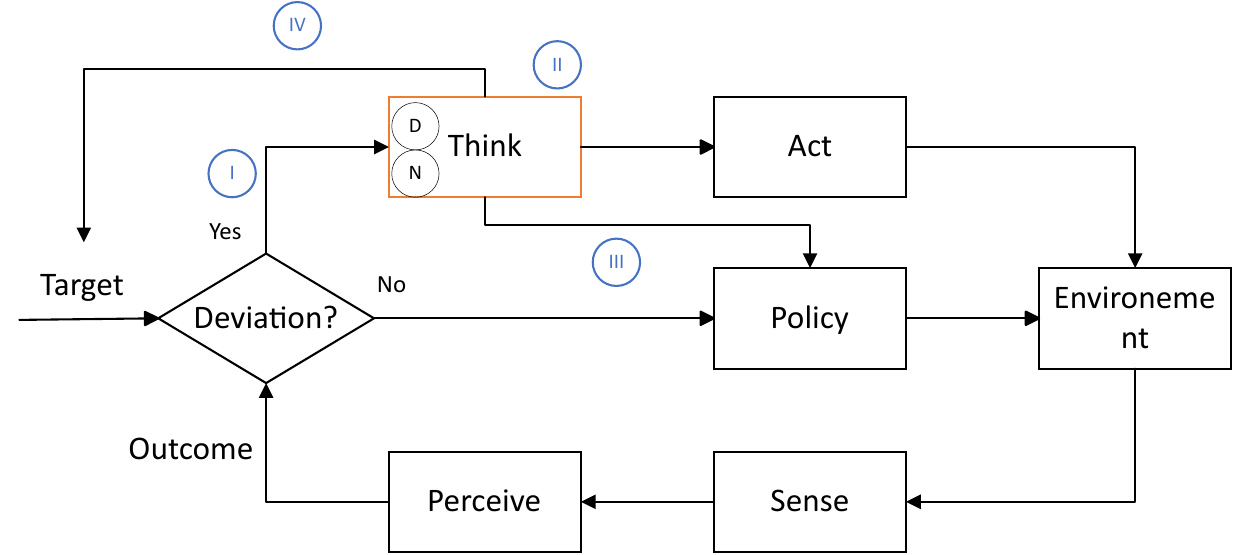} 
\caption{Emotions. \Circled{I} is ``Surprise''. When deviation is worse, \Circled{II} is ``Nervousness''; \Circled{III} alone is ``Relief''; the combination of \Circled{III} + \Circled{IV} is ``Sadness''; the deletion of the control loop is ``Grief''. When deviation is better, \Circled{II} is ``Excitement''; the combination of \Circled{III} + \Circled{IV} is ``Joy''; the creation of a new control loop is ``Ecstasy''; \Circled{IV} alone is ``Disappointment''.}
\label{fig:loop_emotion}
\end{figure}

Emotions are recognized patterns of cognitive operations in the mind. When it receives the signal from the control loop that the outcome deviates from the target, this operation can be recognized as ``Surprise''. When it uses the ``Think-Act'' branch to eliminate the worse deviation, this can be recognized as ``Nervousness''. When a new action is found to restore the original target and it updates the policy, this can be recognized as ``Relief''. When no action is found to restore the original target, one is found to reach a lower target and it updates the policy and target, this can be recognized as ``Sadness''. When no action is found to even reach a lower target and the mind deletes the control loop, this can be recognized as ``Grief''. The emotions are shown in Figure \ref{fig:loop_emotion}.

If the deviation is better, the mind temporarily raises the target to the value of the outcome and attempts to find a new action to reach this higher value. This operation is recognized as ``Excitement''. If a new action is found, it updates the policy and permanently revises the target up. This is recognized as ``Joy''. If a new goal is found, it creates a new control loop for the new goal. This is recognized as ``Ecstasy''. If no new action is found, it reverts to the original target and keeps the policy. This is recognized as ``Disappointment''. The emotions are shown in Figure \ref{fig:loop_emotion}.

In Table \ref{tb:emotions}, I list the emotions and the cognitive operations they correspond to. For the emotions not listed in the table, they are recognized patterns of cognitive operations as well. I'd like to use the example of ``five stages of grief'' to briefly present a few more emotions and illustrate a more nuanced process in the mind. A comprehensive list of emotions and their corresponding cognitive operations is left for future work. 

When the mind receives the signal on a deviation of losing something, it first confirms whether the deviation is true and this manifests as ``Denial''. Once it confirms the deviation is true, it assembles the ``Think-Act'' branch and attempts to restore the target. It appraises the cause of the deviation. If the cause is an external entity, it attempts to find an aggressive action to change the behavior of the entity. This manifests as ``Anger''. If the cause is the mind itself, it attempts to change its own behavior and this manifests as ``Bargaining''. When the mind fails to find an action to restore the target or reach a lower target after it exhausts the learning budget, it deletes the control loop. This manifests as ``Grief'' (This stage is often called ``Depression''). When the mind completes the deletion of the control loop, this manifests as ``Acceptance''. The mind can appraise multiple causes and switch back and forth between ``Anger'' and ``Bargaining'' many times. 

\begin{table}[t]
\centering
\begin{tabular}{lp{0.6\linewidth}}
    \hline
    Emotion & Cognitive operations \\
    \hline
    Surprise & Receive the signal that the outcome deviates from the target \\
    \hline
    Nervousness & Learn a new action when the deviation is worse \\
    Relief & Update the policy when the the target is restored \\
    Sadness & Update the policy and target when a lower target is attained \\
    Grief & Remove the control loop\\
    \hline
    Excitement & Learn a new action when the deviation is better \\
    Joy & Update the policy and target when a higher target is attained \\
    Ecstasy & Create a new control loop \\
    Disappointment & Revert to the original target when a higher target is not attained \\    
    \hline 
\end{tabular}
\caption{Emotions and Corresponding Cognitive operations}
\label{tb:emotions}
\end{table}

\subsection{Complexity in Operating Closed Control Loops}

One complexity comes from the capability of the mind to to adjust all the components of the closed control loop. It can find new actions and adjust the target of the goals, as shown in Section \ref{sec:basics}. It can also adjust the sensing and perception part of the control loop, such as changing the sampling frequency in sensing or tuning the conversion of sensory data into action outcome. It can adjust the sensitivity in detecting deviations by changing the tolerance on the difference between the outcome and the target. It can adjust the intensity of ``Think'' by changing the amount of parallel processing. It can adjust the duration of ``Think'' by changing the number of trails $N$ and $D$. The mind needs to fine tune the components to achieve good effectiveness and efficiency.

Another complexity comes from the interactions among control loops. The mind has many goals. These goals are interconnected, sometimes through hierarchies. A single event may trigger deviations in multiple control loops. An action may change the outcomes of multiple control loops. The mind needs to make complex trade-offs among control loops to decide the action to execute. For instance, when a person wants point out the error of a friend, he may need to trade off between helping the friend realize the error and not making the friend feel embarrassed. 

One more complexity comes from the mind's capability of imagining the future. The mind gains experiences over time. It builds a world model with experiences and acquires the capability of predicting the progression of a goal over time. It sets up intermediate goals and targets for early detection of deviations. When a deviation is detected for an intermediate goal, it predicts the outcomes of alternative actions including no action in its world model and decides the action to take. It experiences emotions as it tries out fictional actions in its world model within its control loops. The mind needs to assess well the probabilities of its prediction. 

\section{Related Literature} \label{sec:literature}

\subsection{Connection with Existing Theories on Emotions}

\begin{figure}[t]
\centering
\includegraphics[width=\columnwidth]{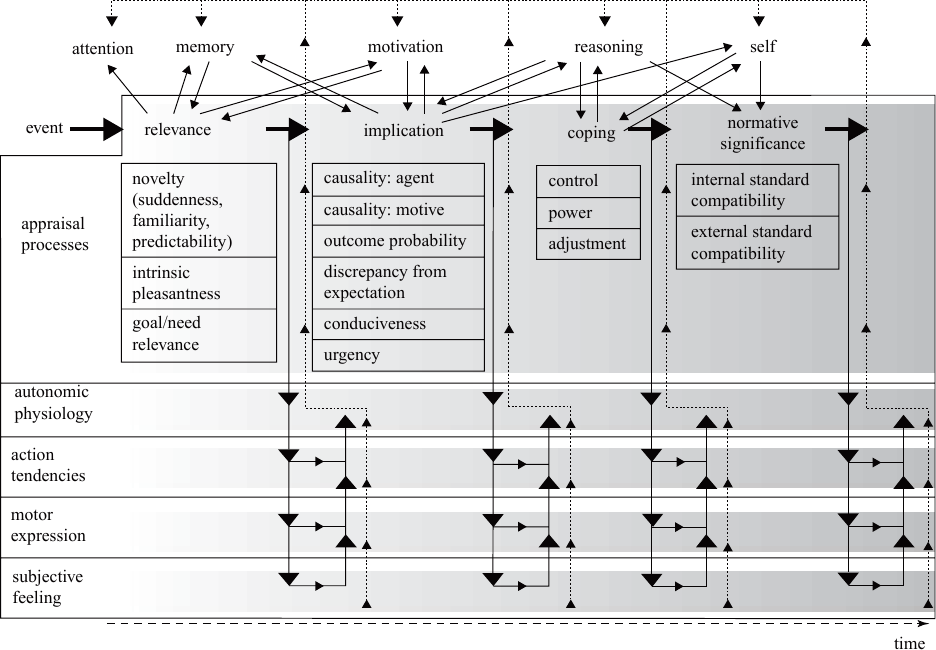} 
\caption{\cite{SANDER2005317} component process model}
\label{fig:scherer}
\end{figure}

My theory is mostly related to appraisal theories of emotions. One of the more recent and impactful appraisal theories is the Scherer’s component process model (CPM) in \cite{SANDER2005317}, as shown in Figure \ref{fig:scherer}. Many elements correspond to each other between my theory and the CPM model. 

In my theory, the mind continuously appraises whether the action outcome conforms to the goal target. This corresponds to ``novelty'' and ``goal / need relevance'' in CPM. It appraises whether the outcome is better or worse than the target. This corresponds to ``intrinsic pleasantness''.

The mind assembles the ``Think-Act'' branch and learns to eliminate the deviation. If it needs to find a new action in a short period of time, it invokes a lot of parallel processing. This corresponds to ``urgency'' in CPM. It needs to appraise the cause of the deviation before devising an action. This corresponds to ``causality: agent'' and ``causality: motive''. It needs to appraise which actions are viable. This corresponds to ``control'', ``power'' and ``adjustment''. The mind predicts the outcome of an action including no action on the (intermediate) goals in its world model. This corresponds to ``discrepancy from expectation'',``outcome probability'' and ``conduciveness''.

``internal standard compatibility'' and ``external standard compatibility'' are goals that are affected by the action the mind uses to eliminate the deviation. The mind needs to make complex trade-offs among multiple goals when deciding an action.

The criteria of CPM naturally arise in the mind's pursuit of goals with closed control loops. They concerns finding a new action. My theory goes beyond this and covers a wider spectrum of cognitive operations in the mind, i.e. updating the policy, adjusting the targets and removing / creating the control loops. 

My theory also has a connection with constructivist emotion theories. Cognitive operations have different effects on the brain. The operations of finding new actions puts extra cognitive load on the brain. More neurons are firing and more biochemical messages are passing. If the mind needs to try out new actions in the environment, it would bear even higher extra load because it needs to process a lot of information in an alien situation.  When the mind concludes its operations of finding new actions, the extra load winds down and the mind makes changes to the control loop. The mind recognizes these cognitive operations as emotions in relation to whether the deviation is better or worse. This is how the mind constructs emotions. The more universal a goal is across the population, the more homogeneous the operations are and the more universal the emotion is; and vice versa.

\subsection{Five Questions of Emotion Psychology}

Reisenzein \cite{10.1093/oxfordhb/9780199942237.013.014} posed five questions of emotion psychology. I have so far provided the answers to 2 questions, namely ``\textit{Q3: What Is an Emotion?}'' and ``\textit{Q1: How Are Emotions Elicited?}''. I'll provide the answers to the remaining 3 question as follows.

\paragraph{\textit{Q2: What Are the Effects of Emotions on Subsequent Cognition and Behavior?}}

Emotions are recognized patterns of cognitive operations. In the short term, these cognitive operations determine the immediate actions the mind takes. In the long run, the accumulation of these cognitive operations determine the goals, targets and actions the mind has. Moreover, the mind pursues the effectiveness and efficiency of theses cognitive operations. It does so by adjusting attributes of these operations such as sensitivity, intensity, duration, etc. In the short term, this manifests as the mind regulating emotions. For instance, the mind realizes that high intensity in the ``Think-Act'' branch sometimes negatively impacts the effectiveness of the cognitive operations and subsequently reduces intensity. This manifests as ``Calm oneself down when too nervous''. These attributes of the operations correspond to personalities in humans. In the long run, when the mind shifts the attributes permanently, this manifests as a change in personality. When the mind loses some of its adaptability, such as unable to adapt a target or goal, psychological problems arise. I present this aspect in more details in Section \ref{sec:problem}. 

\paragraph{\textit{Q4: Where Do the Emotion Mechanisms Come From; to Which Degree Are They Inherited Versus Learned?}}

The closed control loop is an evolution from the homeostatic mechanism. The homeostatic mechanism  is used to regulate changes and maintain steady internal conditions of an life form. It has most of the elements in a control loop, including the target, the policy, ``Sense'' and ``Perceive''. It applies its policy to regulate changes. It adjusts the target when its policy fails to attain the target over extended period of time. It doesn't have the ``Think-Act'' branch to learn brand-new actions. Many biochemical signals may be shared between the homeostatic mechanism and the closed control loop in a life form. 

The closed control loop is inherited. Which goals to create control loops for, what targets to set for the goals, what actions to use for reaching the targets are learned from experiences. Two minds learn different goals, targets and actions from their unique experiences. They experience different emotions from the same event. The closed control loop may be a simple mechanism. But many control loops interacting with each other at different levels and time scales can produce very complex behaviors. 

\paragraph{\textit{Q5: What Are the Neural Structures and Processes Underlying Emotions?}}

The closed control loop and the various cognitive operations are supported by different brain structures. As emotions are recognized patterns of cognitive operations, all these brain structures are involved in the production of emotions. The construction of emotions is supported by the brain structure of perception since the construction is a conversion of sensory data on cognitive operations to the perception of emotions. 

\section{How Psychological Problems May Arise} \label{sec:problem}

When the mind functions properly, it continuously adapts to its circumstance. It creates and removes control loops, increases or decreases the targets, and finds more efficient actions. It adjusts the sensitivity for detecting deviations and the intensity and duration for eliminating deviations. Through these operations, the mind moves towards goals on which it's most effective and efficient. On these goals, the mind detects deviations and invokes cognitive operations to eliminate deviations occasionally and detects no deviation most of the time. In terms of emotions, the mind experiences emotions occasionally and experiences serenity most of the time. 

Problems arise when the mind loses some of its adaptability. One example is that a child can't lower the unrealistic target his parent obliges him to as he gets punished when he can't reach the target. If the mind can't lower the target when it can't find actions to reach that target, the mind needs to continuously eliminate deviations. It experiences emotions constantly. In its attempts to find actions, the mind increases the sensitivity of detecting deviations and the intensity and duration for eliminating deviations. Over time, the mind gradually becomes overly sensitive to deviations, detecting many minute deviations, and invokes intense cognitive operations over long periods of time. In terms of emotions, the mind appears to be easily triggered over certain things and experiences intense emotions over long periods of time. The problem is similar for the scenario where the mind can't remove a control loop when it can't find actions that have stable outcomes. 

The mind may have goals and targets on the intensity of cognitive operations. When the intensity of cognitive operations is above the targets, the mind detects a deviation and invokes cognitive operations to eliminate deviation. It may find the action of suppressing cognitive operations meets its target on the intensity. The mind applies this action. It appears to be depressed. 

For the mind to regain its adaptability, it needs to scrutinize its emotions consciously or mindfully. It needs to identify the goal that causes the constant emotions and reviews it: deciding whether it should have the goal at all, reviewing whether the target of the goal has a reasonable value, and revising how it finds actions to meet the target.

\section{Discussion} \label{sec:discuss}

I propose that the mind pursues goals with closed control loops and emotions are recognized patterns of cognitive operations. The mind uses the control loop to detect and eliminate the deviation between the action outcome and the goal target. It recognizes the patterns of cognitive operations in the control loop as emotions. My theory contains naturally the criteria of the appraisal theories and constructs emotions from the patterns of cognitive operations. It indicates that psychological problems arise when the mind fails to adapt certain aspects of the control loops. 

Emotions provide means for us to observe the inner working of our own minds. Our minds understand and adapt to the world around us through the cognitive operations in the control loops. Emotions enable us to recognize and improve these cognitive operations. Emotions also mark the important events in our lives that shape us in one way or another. They enable us to select the most relevant events to retain in our memory.

In a social context, emotions exposes our cognitive operations to others and transmits essential information. When we express an emotion and others empathize it, they instinctively understand the cognitive operations we are experiencing. For instance, when we express ``Frustration'' and others empathize it, they understand instinctively that we can't find a new action after many trials. They may then offer their cognitive capacity to help us find a new action. 

A more accurate theory of emotions enables deeper understanding of human minds and supports smoother construction of artificial minds. My proposition offers a fresh perspective and advances the building of such a theory. It brings closer a future where human and artificial minds alike run more effectively and more efficiently.

\clearpage
\bibliography{../References/emotions}

\end{document}